\documentclass[
  preprint,             % <-- or use this instead of `reprint`
  showkeys,
  amsmath,amssymb,
  aps,
  prb,
  floatfix,
]{revtex4-2}

\PassOptionsToPackage{backend=biber}{biblatex}
\usepackage[T1]{fontenc}
\usepackage[utf8]{inputenc}
\usepackage{siunitx}
\usepackage{amsmath} 
\usepackage{bm}% bold math
\usepackage{xcolor}
\usepackage[pdfencoding=auto]{hyperref}
\hypersetup{colorlinks=true, citecolor=blue, urlcolor=blue, linkcolor=black}
\usepackage{xr}
\makeatletter
\newcommand*{\addFileDependency}[1]{% argument=file name and extension
  \typeout{(#1)}
  \@addtofilelist{#1}
  \IfFileExists{#1}{}{\typeout{No file #1.}}
}
\makeatother

\newcommand*{\myexternaldocument}[2][]{%
    \externaldocument[#1]{#2}%
    \addFileDependency{#2.tex}%
    \addFileDependency{#2.aux}%
}
\myexternaldocument[supp-]{supplementary}
\myexternaldocument[ex-]{extendeddata}

\usepackage{graphicx} 
\graphicspath{ {DGraphics_main/} }
\usepackage{caption}
\usepackage{subfig}
\usepackage{wrapfig}
\usepackage[export]{adjustbox}

\usepackage{relsize}% makes table references smaller
\usepackage{multirow}
\usepackage{dcolumn}% Align table columns on decimal point
\usepackage{array}

\usepackage[normalem]{ulem}
\newcommand{\subfigref}[2][\figurename~]{#1\ref{#2}}
\newcommand{\subfigrefs}[3][\figurename~]{#1\ref{#2}~and~\ref{#3}}

\newcommand{\figref}[2][\figurename~]{#1\ref{#2}}

\renewcommand{\eqref}[2][Eq.~]{#1(\ref{#2})}

\DeclareSIUnit\angstrom{\protect\text{Å}}

\definecolor{amethyst}{rgb}{0.6, 0.4, 0.8}

\begin{document}

%\title{Phonon driven non-equilibrium pathways to thermal runaway in battery electrodes} 

\title{Phonon driven non-equilibrium triggers for thermal runaway in battery electrodes} 

\author{Harry Mclean}
\affiliation{Department of Physics and Astronomy, University of Exeter, Stocker Road, Exeter, EX4 4QL, United Kingdom}

\author{Francis Huw Davies}
\affiliation{Department of Physics and Astronomy, University of Exeter, Stocker Road, Exeter, EX4 4QL, United Kingdom}

\author{Ned Thaddeus Taylor}
\affiliation{Department of Physics and Astronomy, University of Exeter, Stocker Road, Exeter, EX4 4QL, United Kingdom}

\author{David W. Horsell}
\affiliation{Department of Physics and Astronomy, University of Exeter, Stocker Road, Exeter, EX4 4QL, United Kingdom}

\author{Steven P. Hepplestone}
\affiliation{Department of Physics and Astronomy, University of Exeter, Stocker Road, Exeter, EX4 4QL, United Kingdom}

\date{\today}

\begin{abstract}

Thermal runaway in lithium-ion batteries is governed by the poorly-understood initiation phase~\cite{Murugan2025, Liu2025}, where localised heating introduces instability. 
Here we identify the three key components that trigger thermal runaway, decreases in local conductivity, heat capacity changes, and intercalation heating, which significantly increase temperature gradients that accelerate battery degradation.
Using a multiscale framework that links atomistic phonon calculations with grain-resolved thermal modelling, we identify large thermal gradients across grain boundaries arising from external heating events and intercalation-dependent thermal properties of Li$_x$ZrS$_2$.
The observed changes in thermal conductivity are due to charge redistribution and bond-strength modulation of the host, in contrast to the existing theory of lithium rattler mechanics~\cite{Qian2016}.
Internal heating events driven by intercalation gives rise to local thermal gradients, finite-speed thermal wave interference, and internal thermal fluctuations that generate mechanical strain and sub-grain thermal breakdown.
These results show that the trigger for thermal runaway is controlled by internal grain architecture and composition, as well as the external environment.
Our findings establish materials and electrode design rules for suppressing hotspot formation and improving battery safety during fast charging.
\end{abstract}

% battery operation forms external heating, i.e. Coulomb heating due to Joule heating.
% Intercalation events are internal heating

% Combined with the additional identified features of 
% All these combine to result in mechanical strain and sub-grain thermal breakdown


\maketitle

\section{Introduction}

The rapidly growing need for fast battery charging is driving lithium-ion batteries toward their electrochemical and thermal limits~\cite{Tomaszewska2019}. As charging currents rise, heat generation intensifies, producing steep temperature gradients that drive degradation of the electrolyte or solid–electrolyte interphase (SEI). This can yield reactive by-products and, in extreme cases, trigger catastrophic failure through rampant decomposition (thermal runaway)~\cite{ThermalManagmentOfBatteries, FastBatteryCharging, BatteriesExploding, Chen2021, Jung2017, Hou2020, Parks2023, Zhao2025,Edge2021}. Despite significant progress, the fundamental trade-off between charging rate and operational longevity/safety remains unresolved~\cite{Tai2024, Gharehghani2024}.

At the electrode scale, thermal runaway proceeds through distinct stages. Except for the first stage, these well-characterised stages proceed as: large currents induce intra-grain fracture~\cite{Zhao2010} (where “grains” are the individual crystalline domains within the electrode), driven by the volume/strain changes that accompany lithium intercalation~\cite{Parks2023}. These strain fields cleave grains where lithium first enters the electrode~\cite{Li2020}, undermining the SEI, initiating formation–breakdown cycles~\cite{Jung2017, Hou2020, Parks2023, Zhao2025} that accelerate degradation and ultimately precipitate thermal failure. The earliest microscopic event, the trigger, that initiates this cascade remains unknown~\cite{Murugan2025, Liu2025}.

At the atomic scale, intercalation modifies bonding, introduces disorder, and alters lattice dynamics. Experiments and simulations on intercalated MoS$_2$ (disorder-enhanced scattering)~\cite{MoS2Kinter}, graphite (rattler-mode scattering and phonon-band flattening)~\cite{Qian2016}, LiCoO$_2$ (bond weakening and structural disorder)~\cite{Cho2014}, and phase-separated systems~\cite{Shin2022} demonstrate that intercalation can suppress lattice thermal conductivity by up to an order of magnitude.

To identify the "trigger" of thermal failure, we reduce it into three regimes: the atomic, the sub-granular, and the multi-grain macro scale. To resolve this, one has to first tackle the phonon dynamics at the atomic scale ({\it atomic}), assess how this affects the macro-scale ({\it macro}), before finally evaluating the dynamic interplay between these scales ({\it sub-granular}). 

Here, our approach involves integrating first-principles phonon calculations ({\it atomic}) with time-resolved, concentration-dependent thermal simulations at differing length scales ({\it sub-granular} and {\it macro}). We track the evolution of lattice conductivity and its influence on temperature distributions within active grains. Our focus is on layered transition-metal dichalcogenides (TMDCs) and graphite electrode; model systems that combine structural simplicity with well-defined intercalation pathways. As a representative TMDC electrode (analogous to NMC), we study $\text{Li}_x\text{ZrS}_2$~\cite{ZrS2DFT, ZrS2Exp}, a promising electrode material distinguished by minimal volume change, a smooth voltage profile, and favourable symmetry for accurate phonon modelling~\cite{Price2024}. For comparison and validation, we also examine graphite, the canonical anode material~\cite{Qian2016}. By establishing the atomic anisotropic properties of these materials~\cite{MoS2Kinter, Qian2016, Cho2014}, we can examine factors that localise heat, resulting in hotspots on the macro scale, and how these macroscopic properties influence the sub-granular structure, resulting in critical effects such as transient thermal waves that can result in sub-grain electrode failure. This pathway enables us to examine the direct quantitative link between the atomic scale and the triggers of device-level thermal runaway.

\newpage
\section{Results: {atomic scale} phonon dynamics}

\begin{figure}[h!]
    \centering
    \subfloat[]{\includegraphics[width=0.45\linewidth]{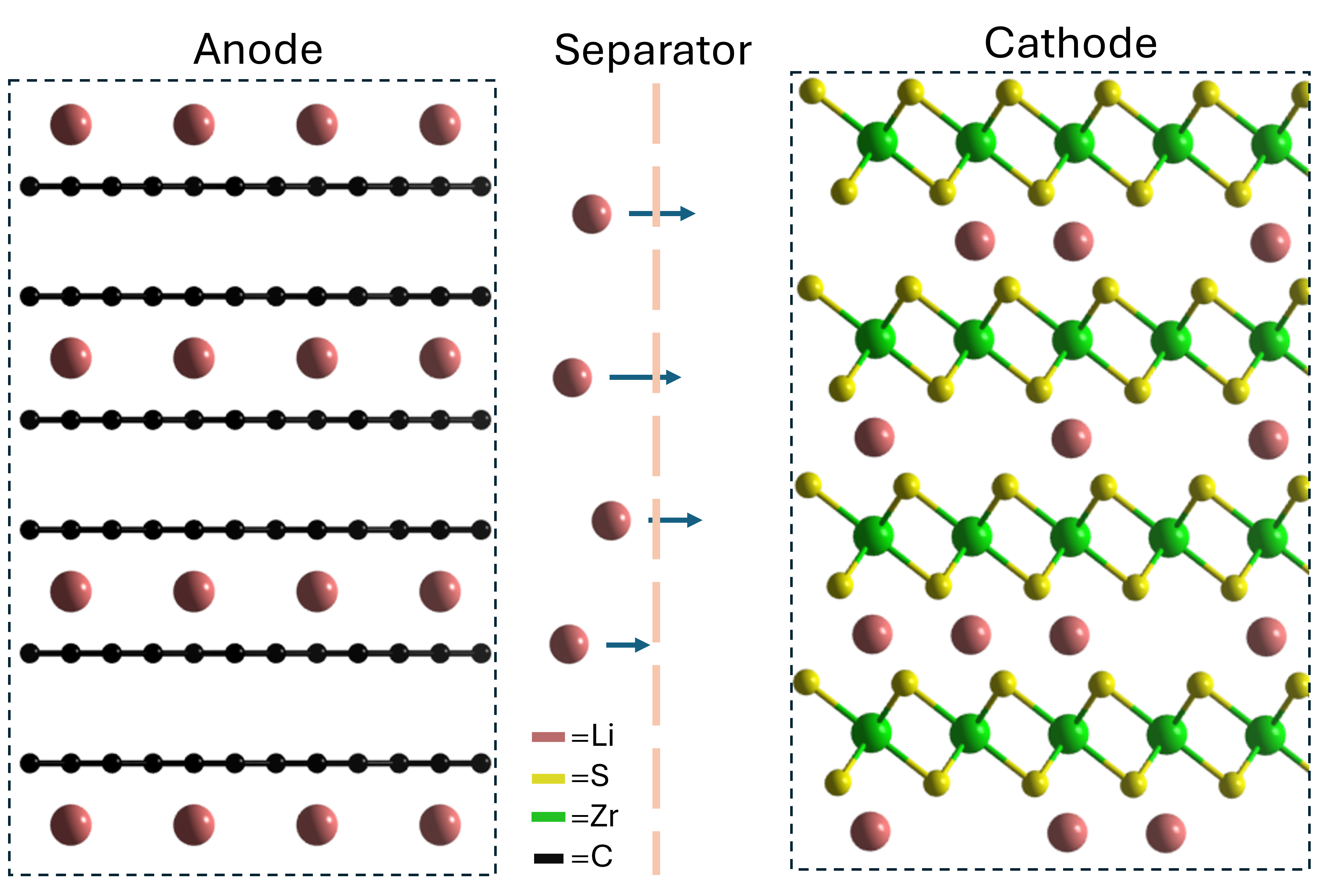}\label{subfig:IntercalationIntoTMDC}}
    \hspace{1em}
    \subfloat[]{\includegraphics[width=0.45\linewidth]{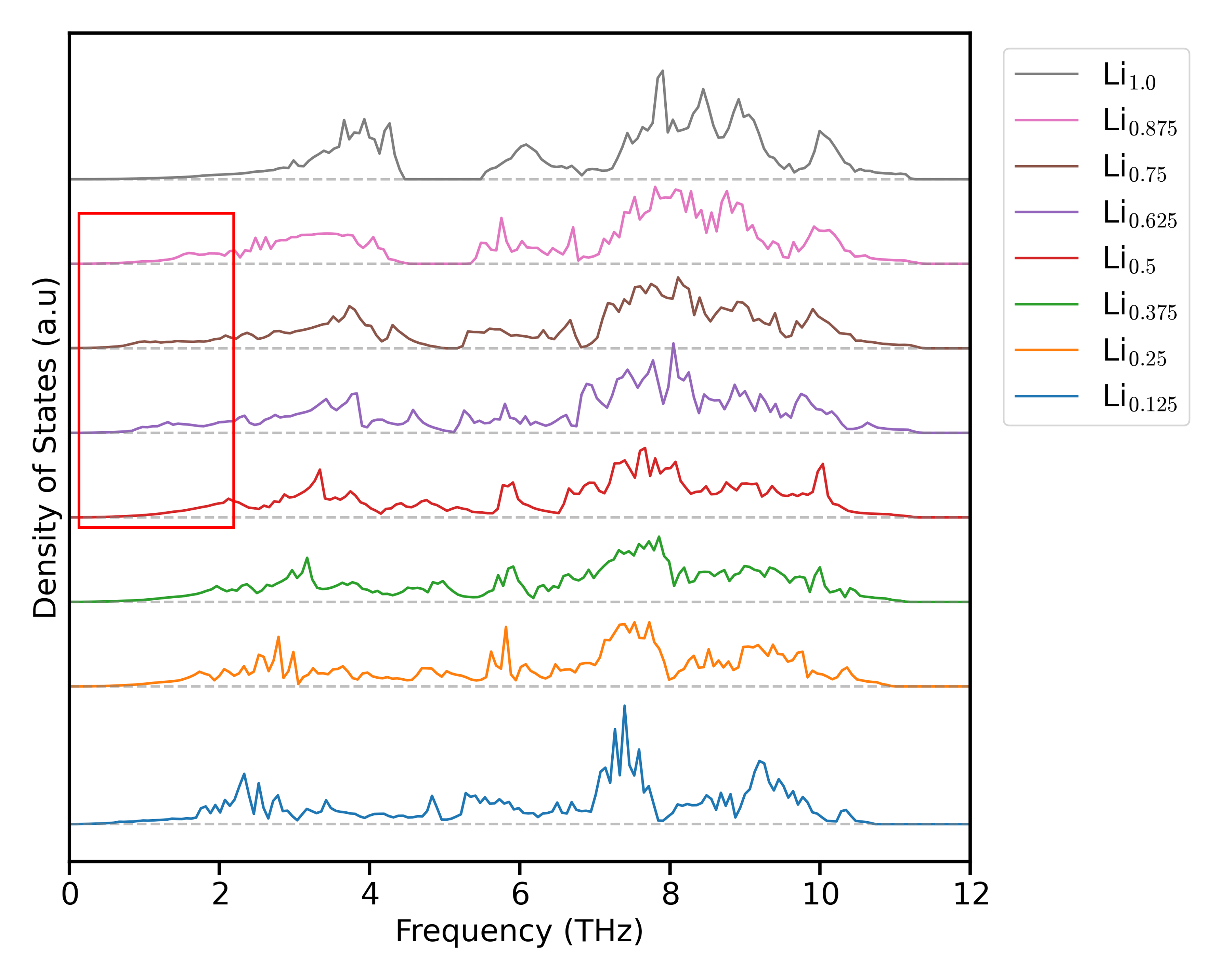}\label{subfig:LiDOS}}
    \hspace{1em}
    \subfloat[]{\includegraphics[width=0.45\linewidth]{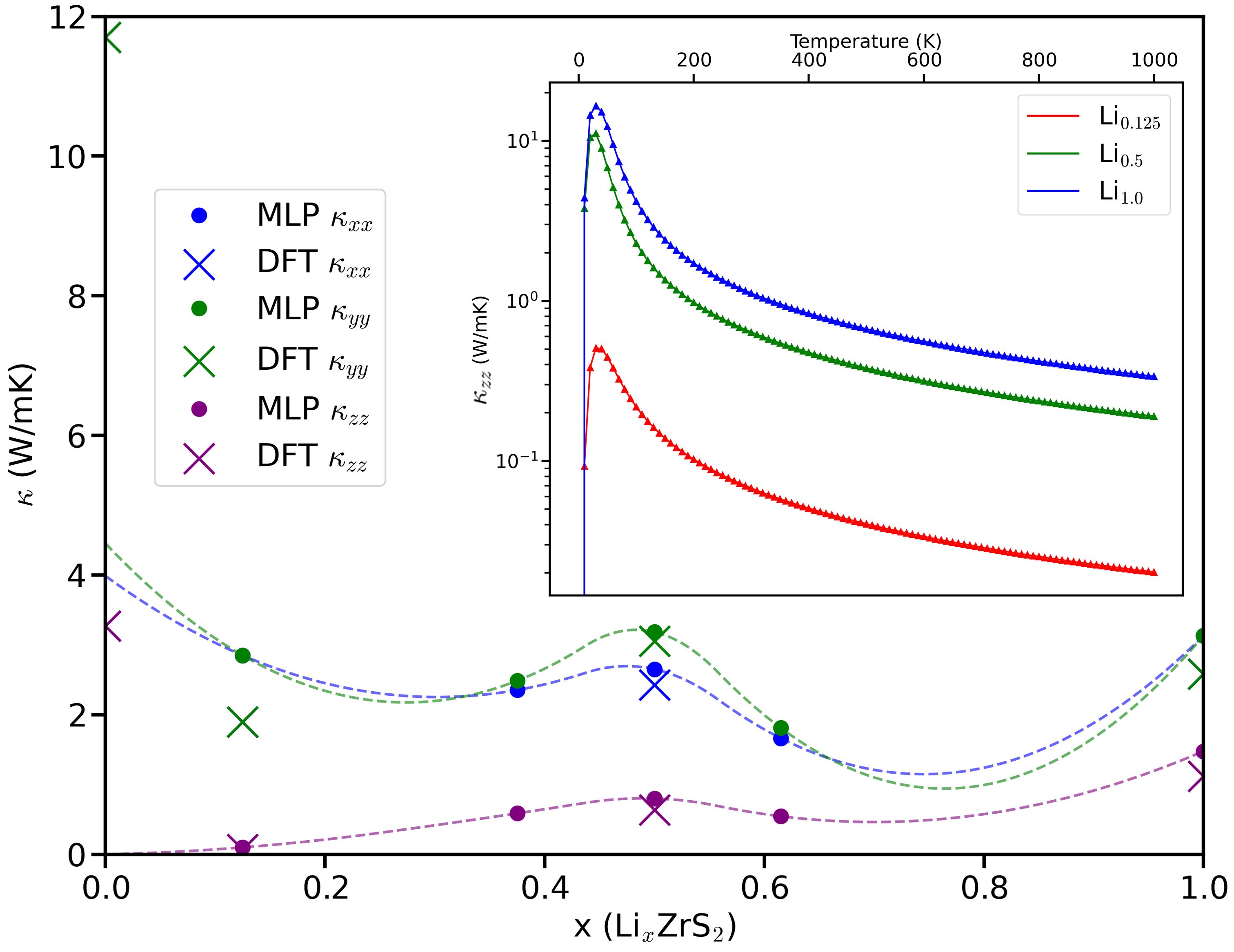}\label{subfig:ThermalConductRTAMLPDFTKTemp}}
    \hspace{1em}
    \subfloat[]{\includegraphics[width=0.45\linewidth]{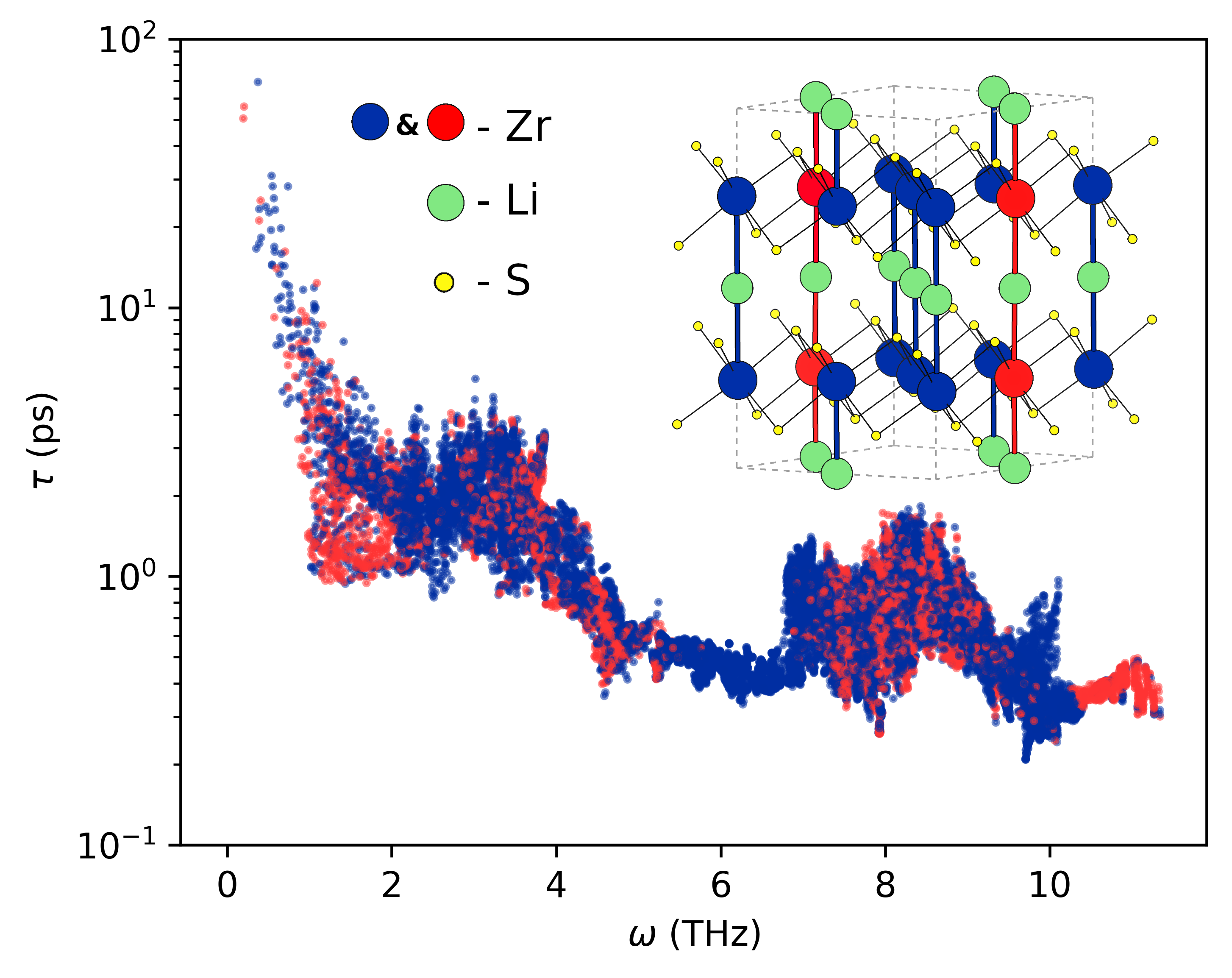}\label{subfig:Li5ZrTau}}
    \caption{\protect\subref{subfig:IntercalationIntoTMDC} Illustration of a discharging lithium ion battery comprised of an anode (graphite), cathode (ZrS$_2$), and a separator.
    \protect\subref{subfig:LiDOS} Phonon density of state for different concentrations ($x$) in Li$_x$ZrS$_2$.
    \protect\subref{subfig:ThermalConductRTAMLPDFTKTemp} Thermal conductivity at 300~K of RTA DFT compared to the RTA MLP technique (see Supplementary data for additional results in %
    Fig.~S7%
    % \suppfigref{subfig:KappaVInt300KCompare}%
    ). 
    Inset of \protect\subref{subfig:ThermalConductRTAMLPDFTKTemp}, the temperature dependence of lattice thermal conductivity (through-plane RTA DFT) for concentrations ($x$) of, $0.125$, $0.5$, and $1.0$. \protect\subref{subfig:Li5ZrTau} The mode-dependent phonon relaxation time at $x=0.625$ (Li$_x$ZrS$_2$) and the contribution from the zirconium to that relaxation time, with the inset showing the arrangement of lithium and ZrS$_2$.}
    \label{fig:1}
\end{figure}

To properly understand the electrode's thermal behaviour during lithium intercalation (\subfigref{subfig:IntercalationIntoTMDC}) we need to assess the phonon properties at the {\it atomic} level. This is necessary, both to explore the underlying mechanisms and, to parametrise the macro-scale transport models based on physical values. The most critical of these, the thermal conductivity, has three contributions determined from the phonons:  heat capacity, phonon velocity and scattering.\\

The heat capacity is derived from the phonon density of states. Upon low-concentration lithiation (from Li$_{0.125}$ to Li$_{0.25}$), a clear 2D-to-3D vibrational transition occurs, evidenced by a change from linear to quadratic low-frequency dependence (\subfigref{subfig:LiDOS}) in the phonon density of states, indicating increased interlayer coupling.  With increasing intercalation, acoustic velocities generally decrease (%
Fig.~S2a%
% \suppfigref{subfig:BANDGroupVel}%
), and a low-lying, dispersive mode emerges at 1~THz (\subfigref{subfig:LiDOS}). This mode softens, becoming flat at Li$_{0.625}$ and forming a sharp phonon density of states peak. This results in: (1) lower acoustic velocities; (2) increased phonon scattering  (reducing thermal conductivity); (3) a phonon bandgap forms, widening to ~1 THz at full lithiation, demonstrating an electrochemically tunable phononic crystal. 

The specific heat capacity of solids at high temperatures is governed by the Dulong Petit limit and increases proportionally to the lithium density.  As such, we can estimate for any electrode the room temperature heat capacity change with lithiation as $\Delta c_v=3x k_B N_A/M_r$ where $x$ is the fraction of intercalation and $M_r$ is the unintercalated molar mass. Hence, as $c_v$ increases linearly with intercalation (%
Fig.~S4b %
% \suppfigref{subfig:HeatCapacityAt300K} %
and %
Fig.~S10g %
% \suppfigref{suppsubfig:HeatCapacityFitGraph} %
for Li$_x$ZrS$_2$ and Graphite, respectively).  This relationship means that as battery electrodes deintercalate, the heat capacity will decrease, resulting in a net temperature increase as a result of conservation of energy.

Graphite and TMDC electrode materials show a general increase in the scattering rate (the inverse of relaxation time) with increasing amounts of lithiation (%
Fig.~S6%
% \suppfigref{fig:PhononRelaxationTime_MLP}%
), arising from a larger phonon scattering space.
This larger phase space is due to a higher number of non-degenerate phonon modes, which leads to a decrease in thermal conductivity.
The ZrS$_2$ thermal conductivities of $\approx$12 W/m/K (agreeing with Glebko et al~\cite{Glebko2019}) decrease with lithiation, as a result of this phenomenon, to between 1.5 and 4 W/m/K for in-plane directions, as shown in ~\subfigref{subfig:ThermalConductRTAMLPDFTKTemp}.

Our thermal conductivity of Li$_x$ZrS$_2$ shows a pronounced dip at $x=0.625$, a common feature in intercalation electrodes \cite{Cho2014, MoS2Kinter, Shin2022, Agne2022}.
While such a drop is often attributed to Li rattler modes \cite{Cahill2014, Qian2016}, our phonon hybridisation ratio calculations show Li-derived vibrations are high-frequency and contribute little at 300~K due to low occupancy (%
Fig.~S1%
% \suppfigref{fig:BANDDOS}%
). Instead, thermal transport is dominated by the heavier electrode atoms (zirconium), ruling out rattling as the primary mechanism. The origin of the dip is driven primarily by the lack of symmetry in the system, resulting in an uneven charge distribution of the intercalating lithium. The asymmetric charge redistribution between zirconium atoms is supported by Bader charge analysis (%
Fig.~S3b%
% \suppfigref{subfig:ZrBaderIndividual}%
),  indicating non-uniform bond stiffening, which modulates phonon scattering rates. We show that the zirconium atoms, which lie between the empty and intercalated lithium regions and have a differing charge from others, show the greatest scattering rate in ~\figref{subfig:Li5ZrTau}. Consequently, this mixed bonding environment means any battery electrode system that has regions where the host material has mixed charge states due to partial intercalation will possess low thermal conductivity.

\clearpage

\section{Results: Macro-scale temperature distribution}

\begin{figure*}[h!]
    \includegraphics[width=\linewidth]{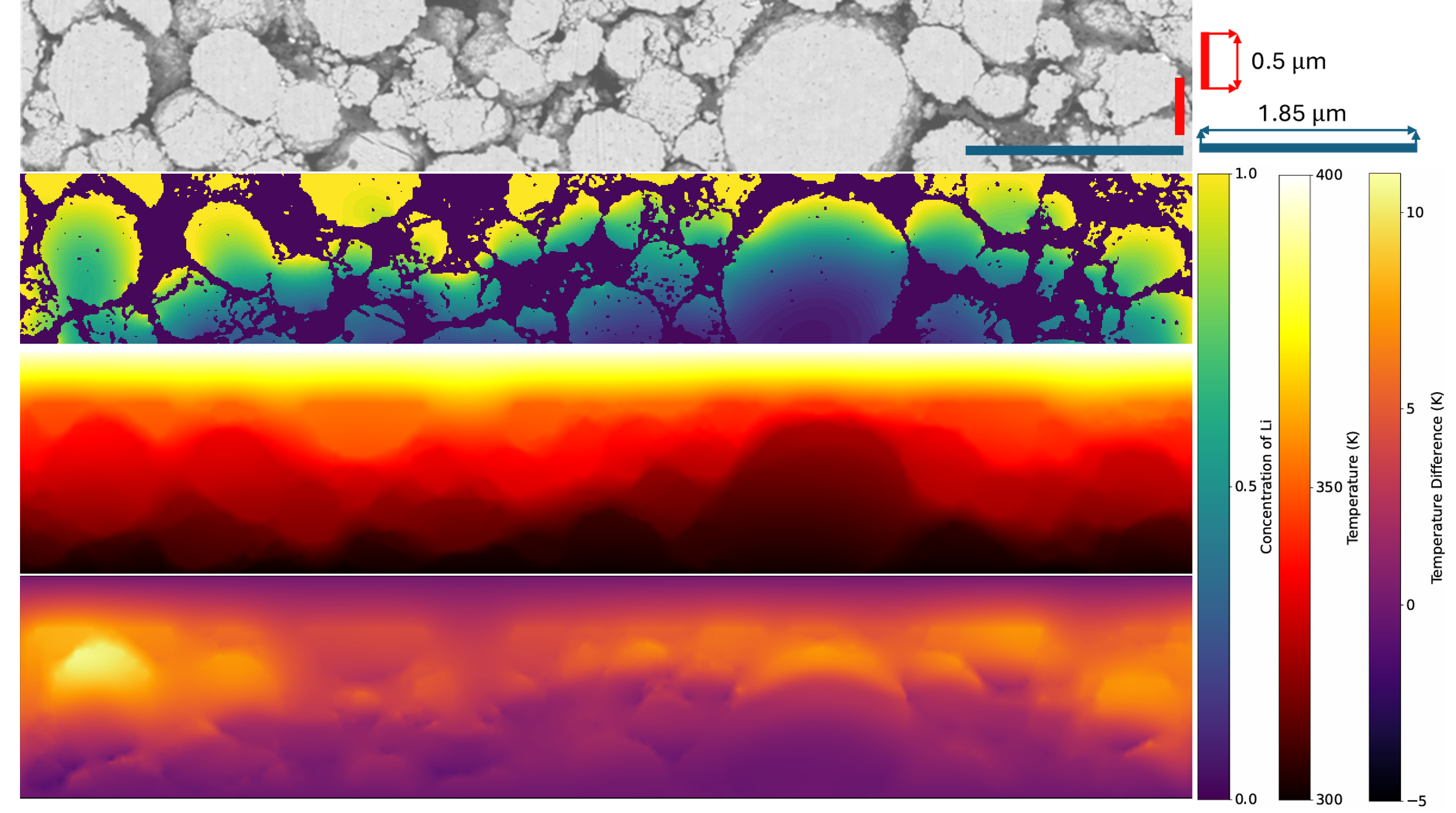}
    \caption{(top) Section of the electrode taken from the experiment~\cite{Cooper2022}. (upper mid) An approximate qualitative distribution of only the ions inside the grains of the electrode shown in (top) at some arbitrary time. (lower mid) The steady-state thermal simulation of the system with concentration-dependent thermal properties. (bottom) The difference in temperature between the concentration-dependent thermal properties and the concentration-independent system, which is based on the pre-charging thermal properties.}
    \label{fig:BatteryExperiment}
\end{figure*}

Having established the atomic scale properties of our electrodes, we now examine the {\it macro} scale ($>$1~$\mu$m) properties of grains and electrodes, exploring how different concentrations of lithium and the resulting dips in thermal conductivity (~\figref{fig:1}) affect the temperature distribution. Several groups~\cite{Xu2020, Lam2024, Daubner2025}  have demonstrated a non-uniform distribution of lithium within grains. Understanding the role of the non-uniform distribution on the thermal performance of the electrodes is crucial when trying to identify the source of thermal runaway.

The three main components of heat generation ($Q$) in a battery cell are resistive (Joule) heating, intercalation heating and residual heating,  $\dot{Q}_J(x, T)$, $\dot{Q}_{\text{int}}(x, T)$ and $\dot{Q}_{res}(x,T)$~\cite{Zhou2022}. $\dot{Q}_J(x, T)$ is the electrode's resistance to current, i.e. "Joule heating" and is an unavoidable part of the heat generation in the charging process. The second component of heating, $\dot{Q}_{\text{int}}(x, T)$, is considered to be associated with the intercalation and de-intercalation of lithium and generates heat in the local vicinity of the lithium intercalation. Other terms, such as side reactions, are challenging to quantify, but result in either local heating similar to $\dot{Q}_{\text{int}}(x, T)$ or broader, more general heating across the overall device, such as the resistive Joule heating and thus understanding  $\dot{Q}_J(x, T)$, and $\dot{Q}_{\text{int}}(x, T)$ captures the thermal behaviour of the electrode.  To tackle Joule heating, we examine a fixed temperature gradient across the system. This approximation effectively models the temperature distribution in grains of thermal heating generated from the battery to circuit connections~\cite{Lyu2023}.

The effect of a non-uniform distribution of thermal properties on an electrode's temperature profile is modelled using a typical granular structure obtained from Cooper \textit{et al.} \cite{Cooper2022} (\figref{fig:BatteryExperiment}). A lattice Boltzmann drift diffusion model reveals a radially decreasing lithium concentration from the grain surface inward, consistent with experimental observations \cite{Xu2020, Lam2024, Daubner2025}. Smaller grains exhibit a more uniform distribution due to shorter diffusion paths. Furthermore, grains closest to the lithium source (top of the figure) have a higher concentration, moderated by grain packing density, which causes closely packed regions near the source to intercalate first.

Using the lithium concentration map, we simulate thermal transport for a ZrS$_2$-like cathode and a graphene-like anode.
Applying a 100~K temperature difference across the cathode simulates an effective Joule heating $Q_J$. The cathode simulation (\figref{fig:BatteryExperiment} lower mid) shows hotspots and visible grain boundaries due to thermal contrasts arising from varying thermal properties. Compared to pre-charging, the temperature difference ranges from -1 to 10.58 K (\figref{fig:BatteryExperiment} bottom). This non-uniformity causes significant internal thermal stress, which is highest in large grains that contain the greatest range of lithium concentrations. The stress permeates the electrode's width, creating a distinct thermal gradient between the separator-facing and current collector-facing sides.  For the high-conductivity anode (%
Fig.~S12%
% \suppfigref{fig:GraphiteElectrode} %
), the pattern is similar, but the fluctuation range is drastically reduced (-0.09 to 0.26 K) due to graphite's (intercalated and unintercalated) much higher thermal conductivity.  Together, these results indicate that for all but the highest thermal conductivities, thermal hotspots will form at grain boundaries throughout the electrode, due to the concentration dependent thermal conductivity.
\clearpage

\section{Results: Sub-grain heating}

\begin{figure*}[h!]
    \centering
    \subfloat[]{\includegraphics[width=0.45\linewidth]{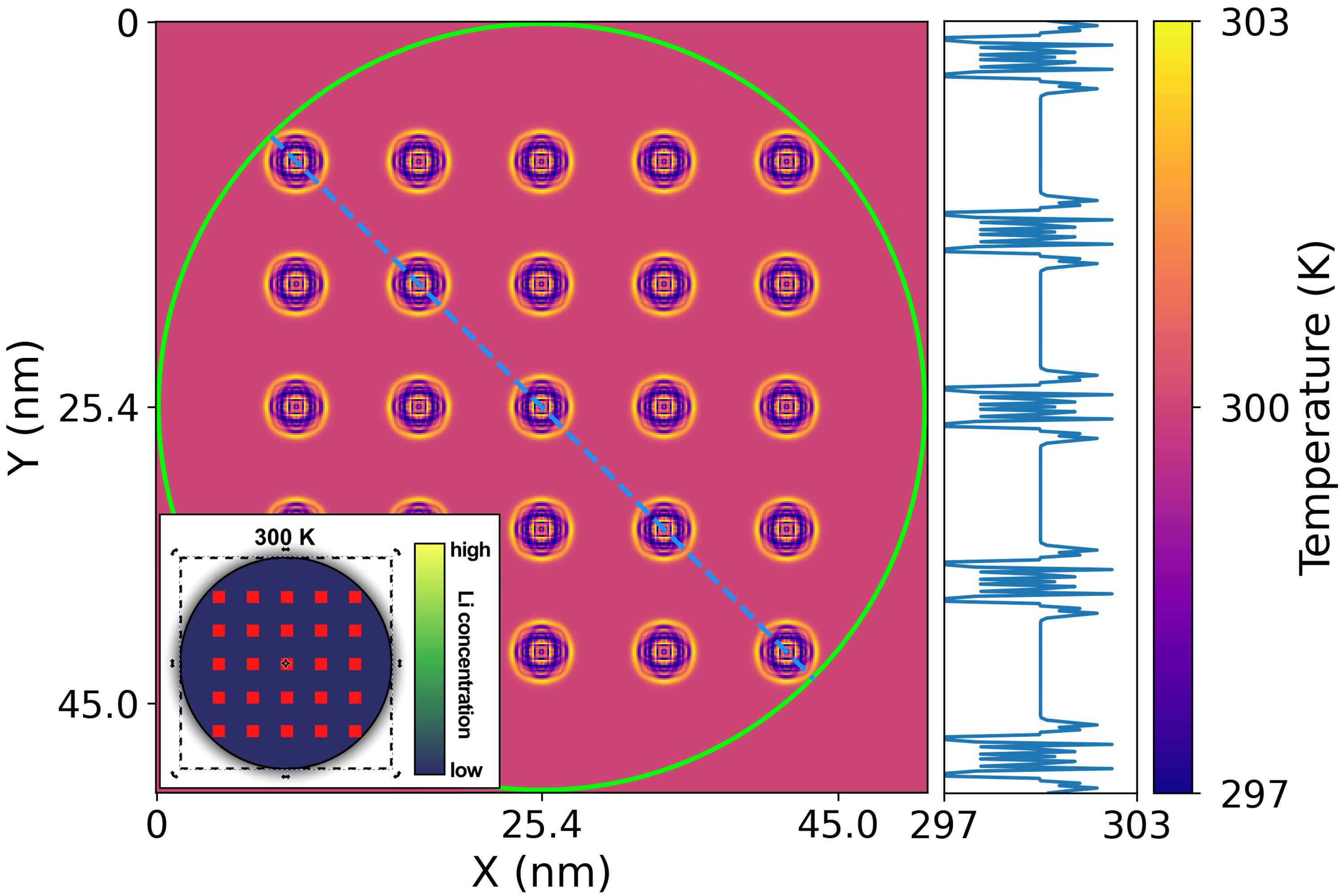}\label{subfig:Uniform3p5ps}}
    \hspace{1em}
    \subfloat[]{\includegraphics[width=0.45\linewidth]{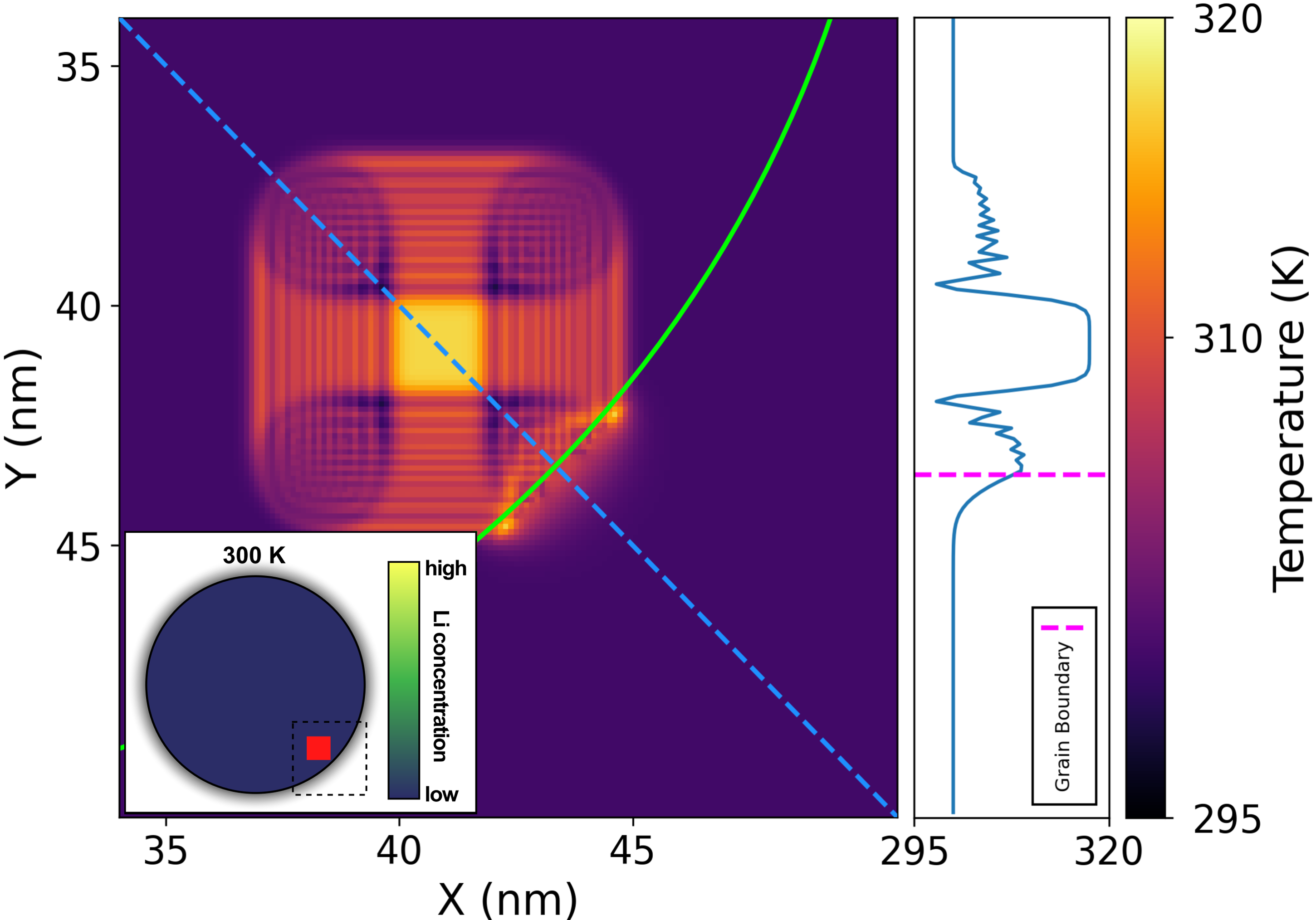}\label{subfig:UniformEdge3p5ps}}
    \hspace{1em}
    \subfloat[]{\includegraphics[width=0.45\linewidth]{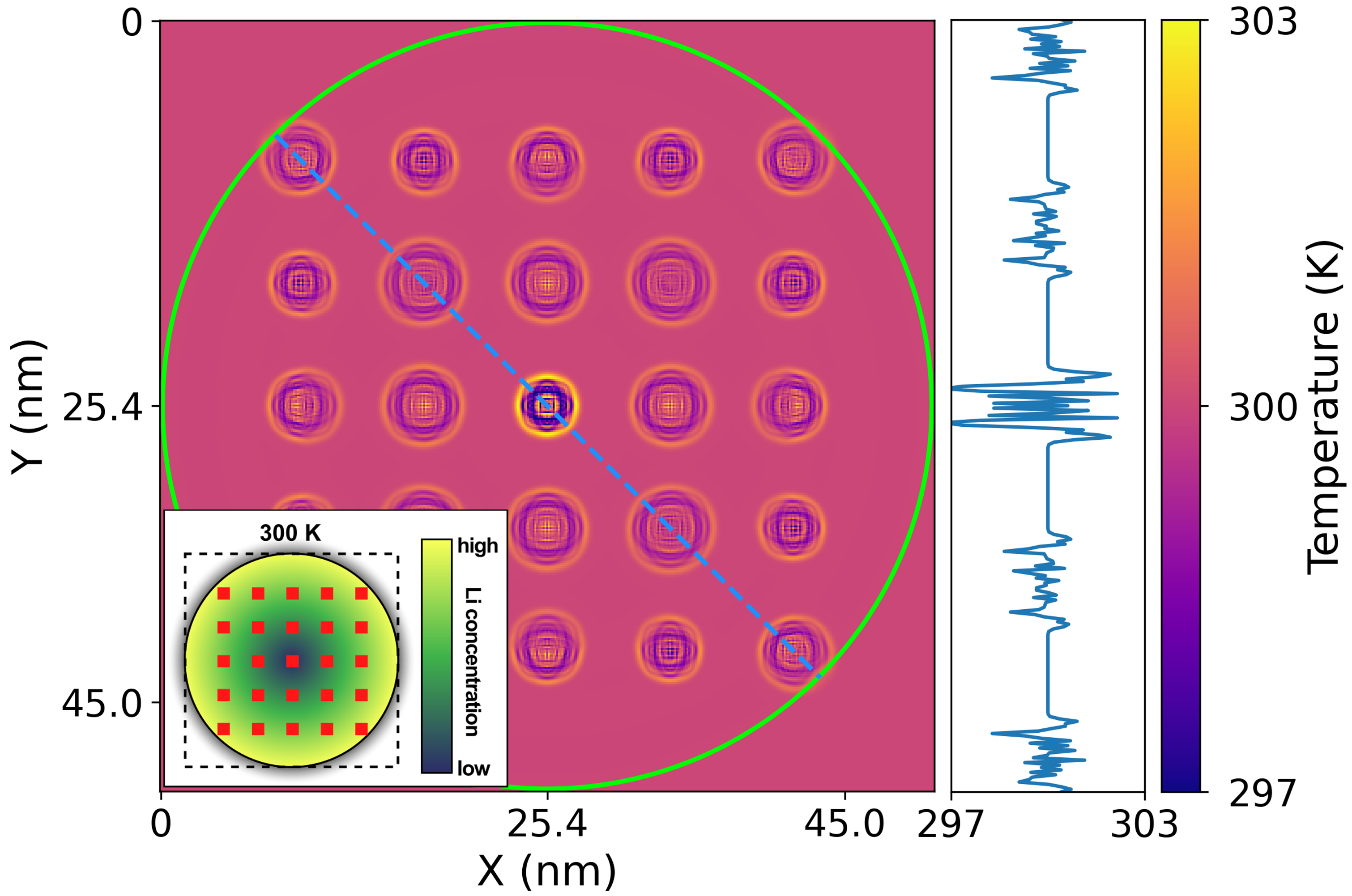}\label{subfig:Graded3p5ps}}
    \hspace{1em}
    \subfloat[]{\includegraphics[width=0.45\linewidth]{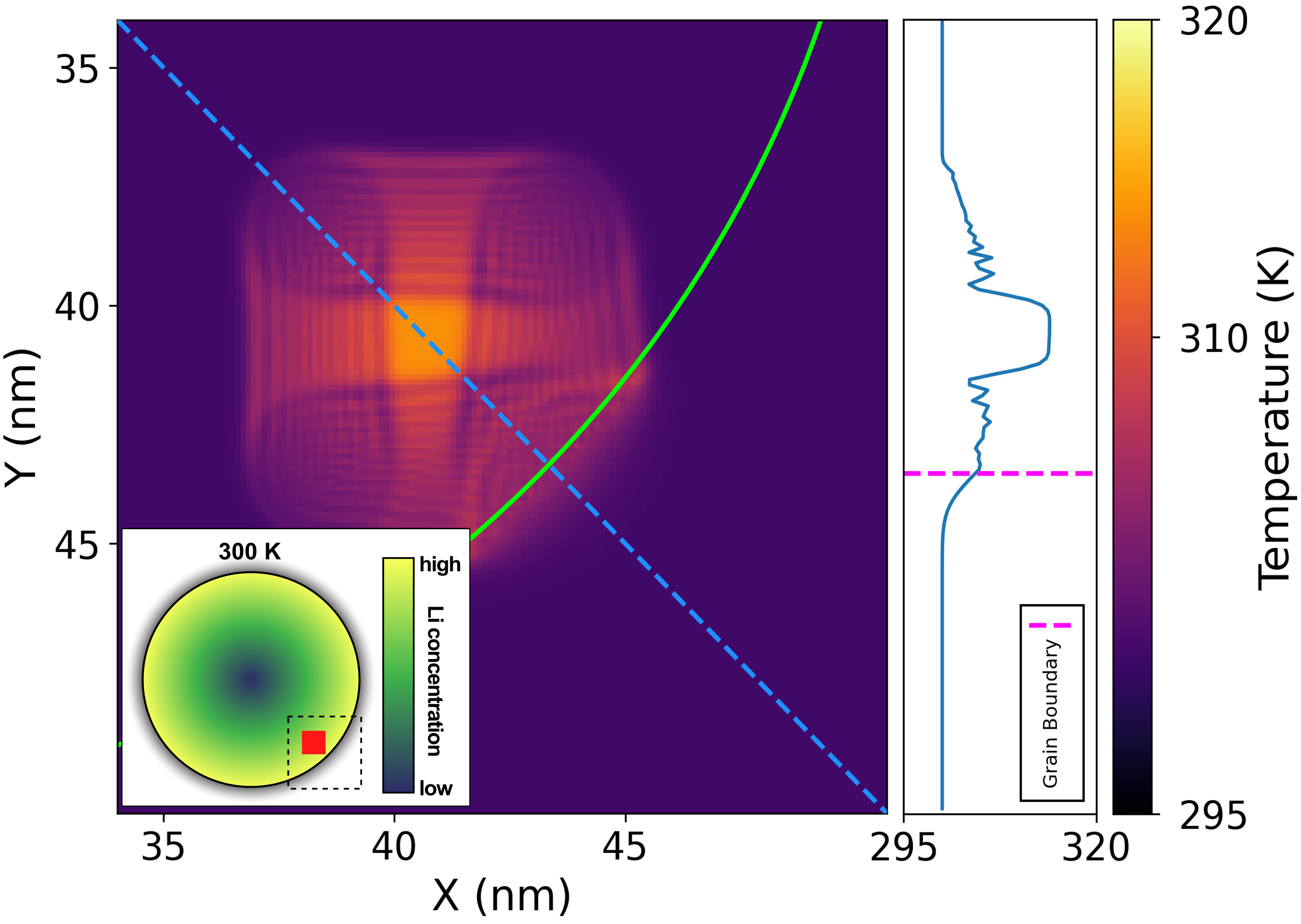}\label{subfig:GradedEdge3p5ps}}
    % \hspace{1em}
    \caption{ \protect\subref{subfig:Uniform3p5ps} and \protect\subref{subfig:Graded3p5ps} show the temperature distribution of the uniform empty and graded grain system, respectively, when the intercalations are distributed evenly throughout the grain centre. \protect\subref{subfig:UniformEdge3p5ps} and \protect\subref{subfig:GradedEdge3p5ps} show the temperature distribution for intercalations with no spacing near the edge of the grain for uniform empty and graded grains, respectively. Figure insets represent schematics of the systems, with the dashed box indicating the view window. All figures show a slice through the 3D system at 3.5~\si{\pico\second}.}
    \label{fig:ThermalGrainsIndividual}
\end{figure*}

Whilst Joule heating dominates the overall thermal management of the system at the macroscopic scale, at the {\it sub-granular} scale, heating due to intercalation is dominated by two factors: (i) the heat generated by each Li intercalated, and (ii) the resulting change in temperature due to the local change in heat capacity.

First, to evaluate the key local heating contribution from lithium intercalation, denoted as $Q_{int}$, we calculate the heat per intercalation event, which is directly related to the entropy change (%
Fig.~S5%
% \suppfigref{suppfig:Entropy}%
). The macroscopic thermal power generated during operation is $P = I \frac{T}{zF} \frac{dS}{dx}$. For Li$_x$ZrS$_2$ (Graphite) at 300~K with $\Delta S \approx 20.3$ J K$^{-1}$ mol$^{-1}$ and a 1 mA current, $P \approx 63$~$\mu$W, ($P \approx 61$~$\mu$W) consistent with experiments \cite{Zhou2022}. 
%For graphite anodes (\suppfigref{suppfig:GraphiteEntropy}), $P \approx 6.1 \times 10^{-5}$ W. 
Scaling to a per-atom basis (where $\sim$6 Li intercalate per fs at 1 mA) yields a conservative heating power of $~10$ $\mu$W per Li for Li$_x$ZrS$_2$.

In addition to the  heating from each intercalation event, we  have to consider that intercalation causes a change in local heat capacity, which, due to energy conservation, must be accompanied by a change in temperature.  As the heat capacity decreases, assuming no significant loss in energy  ($dU = \rho C_V \Delta T=0$) then the temperature must increase (ignoring heat capacity temperature variation). From this assumption (see Methods), we can determine when the concentration locally jumps from $x = 0$ to $x = 1$ (Li$_x$ZrS$_2$), the  change in heat capacity results in temperature changes of up $70$~K (corresponding to a power of $7.9$~$\mu$W), assuming an initial temperature of $T = 300$~K.

By estimating the local power, we evaluate how localised heating events, $Q_{int}$, impact individual grains of electrode material, accounting for both the distribution of intercalation events and variance in thermal properties resulting from the changing lithium concentration profile in the grains. As the time scales of these processes are comparable to those associated with wave-like transient heat transport phenomena, temporal heating effects can be expected to lead to temperature peaks which diffusive models could not account for. To capture this behaviour, we employ our modified form of the Cattaneo heat conduction model~\cite{Mclean2025} (\eqref{eq:FullCattaneoMain}) to simulate the temporal evolution of temperature in individual representative grains.

During cycling, electrode grains transition from a uniform thermal state (pre-intercalation) to one where properties vary radially with lithium concentration, demonstrated in \figref{fig:BatteryExperiment} and %
Fig.~S9%
% \suppfigref{suppfig:validatingLBM}%
. Analysis of spherical grain heating after 3.5 ps (\figref{fig:ThermalGrainsIndividual}) reveals that, despite identical input power, a localised intercalation profile produces a temperature approximately five times higher than uniform intercalation, due to the smaller encapsulated volume.

As heat pulses result in wave-like heat propagation at this timescale, this leads to periodic temperature oscillations under distributed intercalation, visible in \subfigref{subfig:Uniform3p5ps}. In the mid-intercalation case (\subfigref{subfig:Graded3p5ps}), these waves are present, but the system exhibits more varied temperature gradients, driven by local increases in heat capacity between intercalated and non-intercalated regions. This results in a smaller overall temperature range and faster convergence to an isotropic temperature within the grain. The pronounced wave-like behaviour in both systems will amplify mechanical strain and increase the risk of intergranular fracture.

Intercalation of lithium into grains of electrode material rarely occurs uniformly, and hence we can expect to observe more clustered lithium intercalation near the surface of the grain (\figref{fig:BatteryExperiment}). When intercalation is localised near the grain boundary (\subfigrefs{subfig:UniformEdge3p5ps}{subfig:GradedEdge3p5ps}), we see more distinct wavelike behaviour (the over-focusing cross is an unavoidable artefact due to voxel discretisation~\cite{Chai1993}). Critically, at the interface between the grain and the surrounding electrolyte, the temperature is greater than a typical diffuse picture of heating would reveal. Furthermore, as demonstrated by comparing the results from \subfigrefs{subfig:UniformEdge3p5ps}{subfig:GradedEdge3p5ps}, the pre-intercalation profile has an elevated temperature for longer durations, due to the lower heat capacity and higher thermal conductivity of the grain. This prolonged thermal localisation enhances the likelihood of thermally induced degradation, serving as an effective trigger to thermal runaway.

Most electrodes, such as NMC or LiCoO$_2$, have properties very similar to Li$_x$ZrS$_2$.  However,  graphitic carbon, another common electrode, is significantly different from these due to its superior thermal properties.  When comparing the thermal spike in graphite (%
Fig.~S13%
% \suppfigref{fig:GraphThermalGrainsIndividual}%
) with the ZrS$_2$ , we observe the thermal energy is more rapidly dissipated and the magnitude is dampened due to its substantially higher thermal conductivity and heat capacity. For pre- (%
Figs.~S13a and S13b%
% \suppfigrefs{subfig:GraphUniform2ps}{subfig:GraphUniformEdge2ps}%
) and mid- (%
Figs.~S13c and S13d%
% \suppfigrefs{subfig:GraphGraded2ps}{subfig:GraphGradedEdge2ps}%
) intercalation, at 2~ps, the grain exhibits stronger thermal wave behaviour and  interference from reflections.  However, these significant interference patterns would result in high thermal expansion. This strong thermal wave dependence in graphite has been noted~\cite{SSGraphene,Ding2022} by the temperatures in which it is possible to observe second sound.

These results demonstrate the impact of lithium intercalation location and frequency within an electrode grain. They indicate that a key initial triggering factor in thermal runaway is the thermal wave evolution and thermal property distribution inside the grains. Crucially, each time the battery cycles, these factors become more pronounced as lithium clustering increases and grain fractures become more likely. Thus, whilst the first thermal waves caused by intercalation on the first cycle are unlikely to cause any significant effect, the repeated use increases the statistical chance of this breakdown, and in turn accelerates the effect.

%\clearpage

\section{Discussion and conclusion}

Our multi-scale analysis reveals how thermal fluctuations arise and propagate during battery operation.
These form the trigger for thermal runaway.
At the {\it atomic scale}, non-monotonic thermal conductivity stems from the uneven charge distribution on zirconium, which reorganises when lithium occupies symmetry-breaking sites.
The {\it macroscopic scale}, grain-resolved models show that Joule heating and intercalation-driven heat capacity changes create intrinsic temperature deviations of $\approx$10~\si{\kelvin}.
In addition, intercalation events produce finite-speed thermal waves, which amplify interfacial heating and can trigger grain fracture.
This necessitates the need for cooling strategies.
One may consider external cooling methods, such as optimized liquid-cooling layouts \cite{Wang2020}, adaptive airflow \cite{Zhuang2021}, and passive hybrid schemes \cite{Almaasfa2025, Chen2024, Zhao2022}.
However, these external cooling schemes are problematic, as the drastic increase in internal electrode temperatures during fast charging result in large thermal gradients being created.
Traditional modelling, not accounting for dynamic effects (such thermal waves) underestimate these temperatures by up to 50\% at high C-rates \cite{Yang2024}; these factor, in turn, highlight that internal cooling strategies are also needed for improved stability during fast charging.
Our sub-grain simulations reveal that intercalation itself causes the aforementioned localised temperature spikes ($\approx$20~\si{\kelvin}), and indicate adjustments to material-level ({\it subgranular}) strategies, like nanostructured anodes~\cite{Zhou2022}, tuned cathode grains~\cite{Hou2022, Zhao2025}, and optimised solid-state heat pathways~\cite{Agne2022} are needed to improve stability.

Collectively, the three components--decreases in local conductivity, heat capacity changes, and intercalation heating—-generate significant internal strains that accelerate degradation.
Therefore, thermal management must address internal grain architecture alongside macroscopic cooling.
Design rules should prioritise maximising surface-to-volume ratios to improve heat dissipation and stabilise charge redistribution during intercalation.
These approaches offer a direct path to electrodes with greater thermal stability, longer lifespan, and safer fast-charging performance.

\section{Methodology}

First-principles calculations were performed using the VASP within the projector-augmented wave (PAW) formalism and the PBE generalised gradient approximation. Initial atomic structural data and electronic characterisation were obtained from~\cite{Price2024}. A plane-wave cutoff of 700 eV was employed, and spin polarisation was included. Dispersion interactions were accounted for via the DFT-D3 correction. For the $2\times 2\times 1$ supercells used in phonon calculations (phonopy and phono3py), Brillouin-zone integrations employed a $\Gamma$-centred $3\times 3\times 3$ Monkhorst–Pack mesh. SCF convergence criteria were 10$^{-8}$ eV.\\

RTA approximation in the main text is done with a $10\times 10\times 10$ q-mesh (for all intercalations). In addition, an additional $20\times 20\times 20$ q-mesh for a select set of concentrations and Direct LBM ($10\times 10\times 10$ q-mesh) for 0.125, 0.5, and 1.0 intercalations are shown. DFT results and version details  are made available via figshare for reproducibility and use, provided that this work is properly cited. The boundary mean free path is set equal to $1\mu~$m.\\

\subsection{Phonons}
Using DFT (via the use of VASP~\cite{Kresse1993, Kresse1996, Kresse1996_2}) and the Phonopy/Phono3py~\cite{Phonopy} packages. Phonon and thermal properties can be determined by utilising a frozen phonon method for calculating the force constant matrix~\cite{Phonopy}.  Using Phonopy, thermal properties such as heat capacity, entropy, and Helmholtz free energy are obtained. The thermal conductivity is calculated using Phono3py via both the relaxation time approximation and the direct solution to the lattice Boltzmann transport equation. Pypolymlp~\cite{pypolymlp} is an integrated module of Phono3py, which calculates the force constants given a set of random displacements.

\subsection*{Ion Only Numerical Intercalation Code (\textsc{IONIC}) Simulation}

To model the mesoscale distribution of lithium ions within individual grains, we employed a pure drift–diffusion lattice Boltzmann approach implemented in the open-source \textsc{IONIC} code~\cite{ExeQuantCode2025IONICGitHub}. This framework enables efficient simulation of large-scale ionic transport without explicitly resolving electronic or thermal coupling, focusing instead on ion dynamics and concentration evolution at the grain level (~\figref{fig:BatteryExperiment}).  

The \textsc{IONIC} model does not represent a complete battery cell; details of the boundary assumptions and parameter normalisation are provided in the Supplementary Information. Predefined grain geometries were used, and most physical parameters were normalised to unity for clarity and computational efficiency. The relaxation time, $\tau$, was employed to differentiate the distinct material phases, set to $\tau = 1$ for the solid phase and $\tau = 700.5$ for the electrolyte, consistent with literature values~\cite{Jiang2018}.  

Simulations were initialised by positioning an ion-rich region near one boundary, with the system potential subsequently normalised. The model was then evolved in time to establish steady-state charge distributions within the grains. Variation in the initial ion concentration was used to emulate different charging conditions: a larger initial ion population corresponds to a higher effective charging rate. In contrast, smaller populations represent slower intercalation regimes.

\subsection{Thermal modelling}

To model concentration-dependent, time-evolving thermal transport, we employed the Cattaneo-Maxwell-Vernotte (CMV) equation, which generalises Fourier’s law by introducing a finite heat relaxation time, thereby capturing non-Fourier effects relevant at nano length scales and sub-microsecond timescales~\cite{Mclean2025}. The governing equation takes the form

\begin{equation}
\rho C_V\frac{\partial T}{\partial t} - Q
+ \Big[ \rho C_V\tau\frac{\partial^2 T}{\partial t^2}
- \tau\frac{\partial Q}{\partial t}\Big]
= \nabla\cdot(\kappa\nabla T),
\label{eq:FullCattaneoMain}
\end{equation}

where $\rho$ denotes the material density, $C_V$ is the specific heat capacity, $T$ is the local temperature, $t$ is time, $Q$ is the volumetric heat generation rate, $\tau$ is the phonon relaxation time, and $\kappa$ is the thermal conductivity. The terms in square brackets correspond to the hyperbolic (Cattaneo) correction to the classical Fourier model. A detailed derivation of this form from first-principles phonon transport is presented elsewhere~\cite{Mclean2025}.

This continuum-level approach was chosen to capture thermal effects while retaining computational feasibility. Fully atomistic molecular dynamics simulations were deemed unsuitable due to the prohibitive spatial and temporal resolution required to model entire electrode grains.

Thermal simulations were performed using the open-source \textsc{HeatFlow} package~\cite{ExeQuantCode2024HeatFlowGitHub}. For the data shown in \figref{fig:BatteryExperiment}, material parameters correspond to the fitted values in %
Fig.~S10%
% \suppfigref{suppfig:ThermalModellingParams}%
, discretised into 50 uniform intervals. For \figref{fig:ThermalGrainsIndividual}, parameters were discretised into 26 values based on the same fitting dataset.

All systems were simulated in three dimensions with a Dirichlet boundary condition corresponding to a 300~K thermal bath surrounding the model domain. The simulations capture the temporal evolution of local temperature fields during multiple electrochemical intercalation events. The model geometry consists of a 5~$\times$~5~$\times$~5 array (centres equally spaced $\approx$8nm for the distributed system), regions (125 intercalation sites in total), each represented by a 3~$\times$~3~$\times$~3 (9$\times$9$\times$9 for main text high resolution) block of thermally active cells. The edge heating case has the same number of heated sites, with a total of 125 intercalation sites.

%In addition, a continuum approach appears to be valid as work measuring "phonons" across interfaces shows that bulk-like phonon dispersions are recovered at $\approx$~1~nm~\cite{Qi2021}.

\subsection{Deriving effective power of change in heat capacity}
Assuming (for Li$_x$ZrS$_2$) that both density (%
Fig.~S2d%
% \suppfigref{subfig:Density}%
), and heat capacity (%
Fig.~S4b%
% \suppfigref{subfig:HeatCapacityAt300K}%
) are independent of temperature, the change in total internal energy ($dU = \rho C_V \Delta T$) leads to the relation
\begin{equation}
\frac{\rho C_V}{\rho' C_V'} T = T',
\end{equation}
where the primed quantities denote the new properties after intercalation.

If the volumetric heat capacity varies linearly with intercalation fraction $x$ (for Li$_x$ZrS$_2$), we can write $\rho C_V = m x + c$, and the above expression becomes
\begin{equation}
\frac{m x + c}{m (x + \gamma) + c} T
= \frac{\rho C_V}{m \gamma + \rho C_V}T
= \beta T,
\end{equation}
where $\gamma$ represents the change in intercalation fraction.

For a single unit cell where the concentration locally jumps from $x = 0$ to $x = 1$ (Li$_x$ZrS$_2$), we obtain $\beta = 0.768$. This corresponds to a temperature decrease of approximately $70$~K, assuming an initial temperature of $T = 300$~K. If instead you equate the integrals, 
\begin{equation}
    \int^{300}_0 \rho CV dT = \int^{T{\text{new}}}_0\rho'C'_V dT
\end{equation}
Assuming density is temperature independent, $\rho C_V$ is values corresponding to ZrS$_2$, and prime denotes $x=1$ (Li$_x$ZrS$_2$) values. Then you get a temperature change of $\approx 30~K$. This corresponds to a power of $3.4 \times 10^{-6}$~W suggesting the previous method provides an valid order of magnitude estimate.

The associated power can be estimated using
\begin{equation}
\dot{Q} = V \rho C_V \frac{\Delta T}{\Delta t}.
\end{equation}
Using a representative unit cell volume of $V \approx 68 \times 10^{-30}$~m$^3$, we find
\begin{equation}
\dot{Q} = \frac{7.95 \times 10^{-21}}{\Delta t}.
\end{equation}
For illustration, if this process occurs on the order of $\Delta t = 1$~fs, the effective power is approximately $7.9 \times 10^{-6}$~W, which is about $2 \times 10^{-6}$~W lower than the power estimated from intercalation heating ($Q_{\text{int}} \approx 1\times10^{-5}$~W). %
Fig.~S5b %
% \suppfigref{suppfig:BetaVsGamma} %
shows the relationship between $\gamma$ and $\beta$. This gives an estimate on the macroscopic scale for how much lithium has to intercalate at any given instant (before the system has time to thermally equilibrate) before the changing heat capacity/density has a significant effect; it appears to be of the order of 100-1000 moles per m$^3$ before $\beta$ becomes $\ne1$.

\subsection{Computational modelling}
Thermal transport modelling was done with the \textsc{HeatFlow} package~\cite{ExeQuantCode2024HeatFlowGitHub}.
Ion distribution was modelled with the \textsc{IONIC} package~\cite{ExeQuantCode2025IONICGitHub}

\section{Data availability}
Data corresponding to this research, including modelling input files and results, have been made available via figshare (doi:10.6084/m9.figshare.29661095). The code implementations have been made publicly available under the GPLv3 license and can be found on GitHub~\cite{ExeQuantCode2024HeatFlowGitHub, ExeQuantCode2025IONICGitHub}.

\acknowledgements
N.T.T was supported by the Government Office for Science and the Royal Academy of Engineering under the UK Intelligence Community Postdoctoral Research Fellowships scheme for this work (Grant No. ICRF2425-8-148).
We thank the EPSRC for funding H.M (EPSRC-690010152) and F.H.D (EP/X013375/1).
S.P.H was supported by UKRI funding (UKRI2710, EP/X013375/1).
This work used the ARCHER2 UK National Supercomputing Service (https://www.archer2.ac.uk) Via our membership of the UK's HEC Materials Chemistry Consortium, which is funded by EPSRC (EP/X035859).
The authors acknowledge the use of resources provided by the Isambard 3 Tier-2 HPC Facility. Isambard 3 is hosted by the University of Bristol and operated by the GW4 Alliance (https://gw4.ac.uk) and is funded by UK Research and Innovation; and the Engineering and Physical Sciences Research Council [EP/X039137/1].

\section*{CRediT author statement}
H.M contributed to the investigation, analysis, data curation, software, writing-original draft, writing-review and editing, data analysis, and visualisation.
F.H.D contributed data analysis and visualisation.
N.T.T contributed to writing-review and editing, and visualisation, supervision, and funding acquisition.
D.W.H contributed to Writing-Review, and funding acquisition.
S.P.H served as the project supervisor, providing analysis, writing-review and editing, project management, and funding acquisition.

\bibliographystyle{unsrt}
\bibliography{references}

\end{document}